\documentclass[a4paper, 11pt]{article} 

\usepackage{graphicx}
\usepackage{wrapfig}
\usepackage{lscape}
\usepackage{rotating}
\usepackage{epstopdf}
\usepackage{authblk}
\usepackage{units}
\usepackage{amsmath}
\usepackage[colorinlistoftodos]{todonotes}

\usepackage{setspace} 
\usepackage[right]{lineno}
\usepackage[margin=2.0cm]{geometry}
\usepackage{multicol}
\usepackage{siunitx}
\usepackage{ulem}

\usepackage{subfig}
\usepackage{float}
\usepackage[skip=2pt, font=small]{caption}

\usepackage[colorlinks=true,linkcolor=blue,urlcolor=blue,citecolor=blue]{hyperref}
\usepackage[sort&compress,super,comma]{natbib}
\usepackage{url,doi}

\usepackage{soul}										

\begin{document}

\setlength\abovedisplayskip{2pt}
\setlength\belowdisplayskip{3pt}

\title{Surface production of negative ions from pulse-biased nitrogen doped diamond within a low-pressure deuterium plasma}


\author[1*]{Gregory J. Smith}
\author[2]{Lenny Tahri}
\author[3]{Jocelyn Achard}
\author[3]{Riadh Issaoui}
\author[1,4]{Timo Gans}
\author[1]{James P. Dedrick}
\author[2]{Gilles Cartry}

\affil[1]{York Plasma Institute, Department of Physics, University of York, Heslington, York, YO10 5DD, UK}
\affil[2]{Aix-Marseille Universit\'e / CNRS, PIIM, UMR 6633, Centre de St J\'er\^ome, case 241, 13397 Marseille Cedex 20, France}
\affil[3]{LSPM-CNRS, UPR 3407, Universit\'e Sorbonne Paris Nord, Avenue JB Clement, 93430 Villetaneuse, France}
\affil[4]{School of Physical Sciences \& National Centre for Plasma Science and Technology (NCPST), Dublin City University, Dublin 9, Ireland }
\affil[*]{E-mail: \href{mailto:gjs507@york.ac.uk}{gjs507@york.ac.uk} }

\maketitle

\begin{abstract}

The production of negative ions is of significant interest for applications including mass spectrometry, materials surface processing, and neutral beam injection for magnetic confined fusion. Neutral beam injection sources maximise negative ion production through the use of surface production processes and low work function metals, which introduce complex engineering. Investigating materials and techniques to avoid the use of low work function metals is of interest to broaden the application of negative ion sources and simplify future devices. In this study, we use pulsed sample biasing to investigate the surface production of negative ions from nitrogen doped diamond.  The use of a pulsed bias allows for the study of insulating samples in a preserved surface state at temperatures between 150$^{\circ}$C and 700$^{\circ}$C in a 2~Pa, 130~W, (n\textsubscript{e}~$\sim$~10\textsuperscript{9}~cm\textsuperscript{-3}, T\textsubscript{e}~$\sim$~0.6~eV) inductively coupled deuterium plasma. The negative ion yield during the application of a pulsed negative bias is measured using a mass spectrometer and found to be approximately 20\% higher for nitrogen doped diamond compared to non-doped diamond. It is also shown that the pulsed sample bias has a lower peak negative ion yield compared to a continuous sample bias, which suggests that the formation of an optimum ratio of defects on its surface can be favourable for negative ion production.

\end{abstract}
\doublespacing

\section{\Large{Introduction}}

Negative ions play an important role in applications including particle acceleration\cite{Ueno2010, Peters2000, Moehs2005, Lettry2014, Faircloth2018}, neutron generation\cite{Welton2016, Antolak2016}, mass spectrometry \cite{Alton1994, Middleton1974, Calcagnile2005, Yoneda2004}, spacecraft propulsion \cite{Rafalskyi2016, Lafleur2015a, Aanesland2015}, microprocessor manufacturing\cite{Vozniy2009a} and neutral beam heating for magnetic confinement fusion (MCF) \cite{Hemsworth2017, Hemsworth2009, Hemsworth2005, Fantz20127}. 

In plasma-based sources, negative ions are produced using either volume production or surface-production processes, however surface production processes are typically dominant where a high current of negative ions are required\cite{Yoneda2004, Hemsworth2017, Ueno2010}. Increasing the negative ion production from surfaces is typically achieved through the introduction of low work function metals such as caesium, which introduces complex engineering challenges\cite{Hemsworth2017,Cartry12017, Hemsworth2020, Fantz20127, Bacal2020}.

Alternative materials to caesium are under investigation, and these include highly oriented pyrolitic graphite (HOPG)\cite{Ahmad2014, Schiesko2009, Schiesko2010c, Cartry12017}, novel electrides \cite{Sasao}, LaB\textsubscript{6}, MoLa\cite{Kurutz2017}, as well as those that have dielectric properties such as diamond-like-carbon (DLC) and diamond\cite{Kumar2011, Schiesko2009, Schiesko2010c, Kogut2019, Achkasov2019, Ahmad2014, Kogut2017a, Dubois2016, Smith2020}.

Dielectric materials are of particular interest for negative ion surface production due to the band gap that exists between their conduction and valence bands. The valence band in dielectrics is typically located at a lower level than many other materials such as metals and semi conductors. However, through the combination of an image potential downshifting the affinity level of an approaching particle and a reduction in electron detachment due to the presence of the band gap, a dielectric material can be used to enhance negative ion surface production\cite{Borisov2000a, J.Los1990, Borisov1996}.

Diamond is of particular interest due to its large band gap (5.5~eV)\cite{Diederich1998} and because of the capability to manufacture it with particular physical properties, such as with a dominant grain size or crystal orientation~\cite{Diederich1998}. It can also be doped to further influence its properties and electronic band structure~\cite{Liu2017a, Tachibana2001, Scholze1996, Baranauskas1999}. Firstly, as a method to increase conductivity at low temperatures through the addition of boron (\textit{p}-type doping)~\cite{Kumar2011, Srikanth2011}, and secondly, to enhance negative ion yield through the use of nitrogen doping (\textit{n}-type doping)~\cite{Smith2020}. 

Although the introduction of nitrogen doping has been shown to enhance the negative ion yield from diamond, it also introduces a strong temperature dependency due to poor electrical conductivity at temperatures below 450$^{\circ}$C~\cite{Smith2020}. Additionally, the magnitude of a continuous negative bias has been shown to influence the negative ion yield\cite{Kogut2019}. This has previously been attributed to the formation of defects on the surface which alters its electronic properties\cite{Ahmad2014, Kogut2019, Achkasov2019}. It is therefore of significant interest to investigate the negative ion yield from micro crystalline nitrogen doped diamond (MCNDD) with a preserved sample surface and at temperatures where the material is non-conductive.

In this study, we investigate negative ion production using a pulsed sample biasing technique. The use of a pulsed bias lowers the positive ion average energy which preserves the surface state of the sample, and enables the surface biasing of an insulating surface. The negative ion yield is compared between un-doped micro crystalline diamond (MCD) and MCNDD films. This comparison is carried out over a temperature range between 150$^{\circ}$C to 700$^{\circ}$C to develop the understanding of negative ion production from diamond at temperatures below 450$^{\circ}$C where MCNDD is non-conductive\cite{Baranauskas1999, Al-Riyami2010, Bhattacharyya2001}. The experimental method is described in section~\ref{sec:exp_set_and_method} and the results are presented in section~\ref{sec:pulsedbiascomparison}. 

\section{\Large{Experimental setup and method}}\label{sec:exp_set_and_method}

\begin{figure}[H]
	\centering
		\includegraphics[]{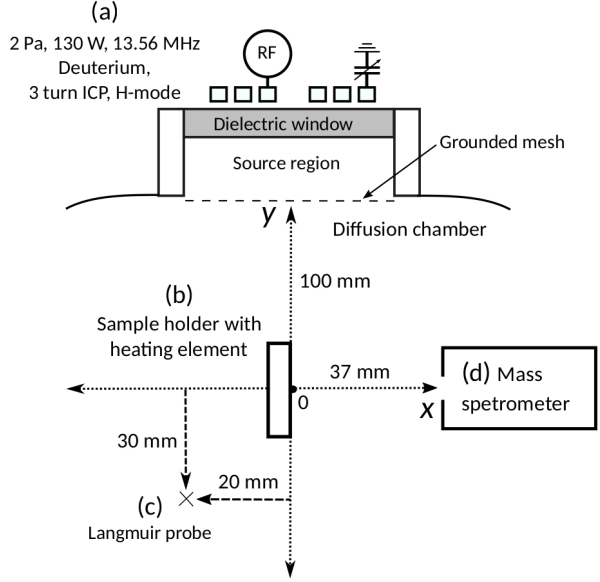}
		\caption{Schematic of the experimental setup.
		(a) Plasma source with associated experimental settings.
		(b) Sample holder and mass spectrometer positioned within diffusion chamber.
		(c) Langmuir probe positioned behind the sample holder. (d) Position of the mass spectrometer relative to the sample holder and Langmuir probe. The schematic is not to scale.}
		\label{fig:Experimental_diagram}
\end{figure}

The experimental setup is shown in figure \ref{fig:Experimental_diagram}.

\subsection{\normalsize{Plasma source and diffusion chamber}}\label{sec:plasma_source_characterisation}

The reactor consists of a cylindrical plasma source (100~mm height and 150~mm diameter), which is separated from a spherical diffusion chamber (200~mm diameter) by a grounded mesh (285~\si{\micro\meter} spacing, hole size 200~\si{\micro\meter}). The mesh reduces radio-frequency (RF) plasma potential fluctuations in the diffusion chamber, which would otherwise alter the shape of the negative ion energy distribution functions (NIEDFs) measured by the mass spectrometer\cite{Ahmad2014}. The mesh also acts to increase the confinement of the plasma to the source region, thereby reducing the plasma density in the diffusion region where the sample is located. The stability of the plasma is therefore tracked in the diffusion region, described in section~\ref{sec:langmuir_probe_motivation} and \ref{sec:mass_spec}, as distinct from the source region\cite{Smith2020}.

A low-pressure inductively coupled deuterium plasma was generated using an RF power generator (Huttinger~PFG~1600~RF) attached to a matchbox (Huttinger~PFM~3000~A). Power is coupled to the plasma through a 3 turn copper coil positioned on top of a dielectric ceramic window (150~mm diameter). The effective power coupled to the plasma was 130~W as measured by the generator. 

The pressure in the diffusion chamber was maintained at 2~Pa, as measured with a Baratron gauge (MKS), using a mass flow controller (7.6~sccm, BROOKS 5850TR) and a 150~mm diameter Riber gate valve installed in front of a turbo molecular pump (Alcatel ATP400). These experimental conditions were selected to reduce the ion bombardment of the samples between application of negative bias and so limit their degradation during the course of the experiment.

\subsection{\small{Sample temperature control}}\label{sec:sample_holder}

Figure~\ref{fig:Experimental_diagram}~(b) shows a schematic of the temperature controlled sample holder. This was described in detail in Ref.~\citenum{Smith2020} and therefore an overview is provided here. 

The sample holder was attached to a 4-axis manipulation arm, which enables the positioning of the samples within the diffusion chamber and alignment of the surface to the orifice of the mass spectrometer. To maximise the number of negative ions collected by the mass spectrometer, samples were positioned at a distance of 37~mm away from its orifice. Previous work has shown that this distance has a negligible effect on the shape of an NIEDF measured by the mass spectrometer~\cite{AAhmad12013}. The alignment of the samples was regularly checked by rotating the sample and maximising the negative ion signal.

A tungsten element, which is controlled by a proportional–integral–derivative controller (PID) connected to a thermocouple installed into the frame of the sample holder, heats the back of the samples. The temperature of the sample surface was determined by comparing the temperature measured using the PID with the temperature measured by a thermocouple attached to the front of a calibration sample. This is expected to introduce an uncertainty of $\pm$20$^{\circ}$C for all temperature measurements.  

\subsection{\small{Deuterium plasma characteristics}}\label{sec:langmuir_probe_motivation}

Measurements of the deuterium plasma potential, electron temperature and density were made using a Langmuir probe (Smart probe from Scientific Systems)\cite{Hopkins1995} within the diffusion chamber as shown in figure~\ref{fig:Experimental_diagram}~(c). It was not possible in this experimental campaign to position the Langmuir probe in front of the sample, meaning measurements were made in a position close to the centre of the diffusion chamber, which gives an indication of the plasma properties within the chamber, as shown in figure~\ref{fig:Experimental_diagram}. 

The Langmuir probe was RF compensated, with a cylindrical tungsten probe tip of length 6.5~mm and 110~\si{\micro\meter} radius. The tip was cleaned prior to each measurement by biasing it with a high positive voltage to heat the probe tip and vaporise any impurities.  Representative values of the electron density, (2.5~$\pm$~0.5)$\times$10\textsuperscript{9}~cm\textsuperscript{-3}, electron temperature, (0.6~$\pm$~0.5)~eV, plasma potential (2.6~$\pm$~0.1)~V and floating potential (0~$\pm$~0.1)~V have been determined directly from the \textit{I(V)} curve obtained with the Langmuir probe. From these values the sheath width adjacent to the sample surface can be estimated and is roughly 1~mm and 2~mm for sample biases of -20~V and -130~V, respectively\cite{Lieberman2005}.   

\subsection{\small{Micro-crystalline diamond samples}}\label{sec:MCDMCBDD}

Micro crystalline diamond (MCD) films were prepared using plasma enhanced chemical vapour deposition (PECVD) as described in Ref.~\citenum{Silva2009}. Nitrogen doped diamond films were produced using a similar PECVD technique to the MCD samples\cite{Smith2020}. The PECVD process utilised a bell jar reactor with a pressure of 200~mbar, microwave power at 3~kW, substrate temperature of 850$^{\circ}$C, background hydrogen gas mixture with a methane concentration of 5\%. The ratio of nitrogen in the gas mixture was set as a means to vary the concentration of nitrogen in the MCNDD film. Each film was deposited on to a $($100$)$ oriented silicon wafer.

In previous work, the nitrogen introduced in the gas phase has been correlated to the nitrogen content in the samples via Raman spectroscopy\cite{Smith2020}. The introduction of the nitrogen in the gas phase was also shown to affect the structure of the crystals grown, changing both the size and the orientation of the crystals. This is described in additional detail in previous work\cite{Smith2020}.

The gas phase content of the MCNDD sample used in this study was 50~ppm. In previous work, it was observed that this concentration of doping resulted in the highest negative ion yield\cite{Smith2020}.

\subsection{\small{Measurement of negative ion energy distribution functions}}\label{sec:mass_spec}

\begin{figure}[H]
	\centering
		\includegraphics[]{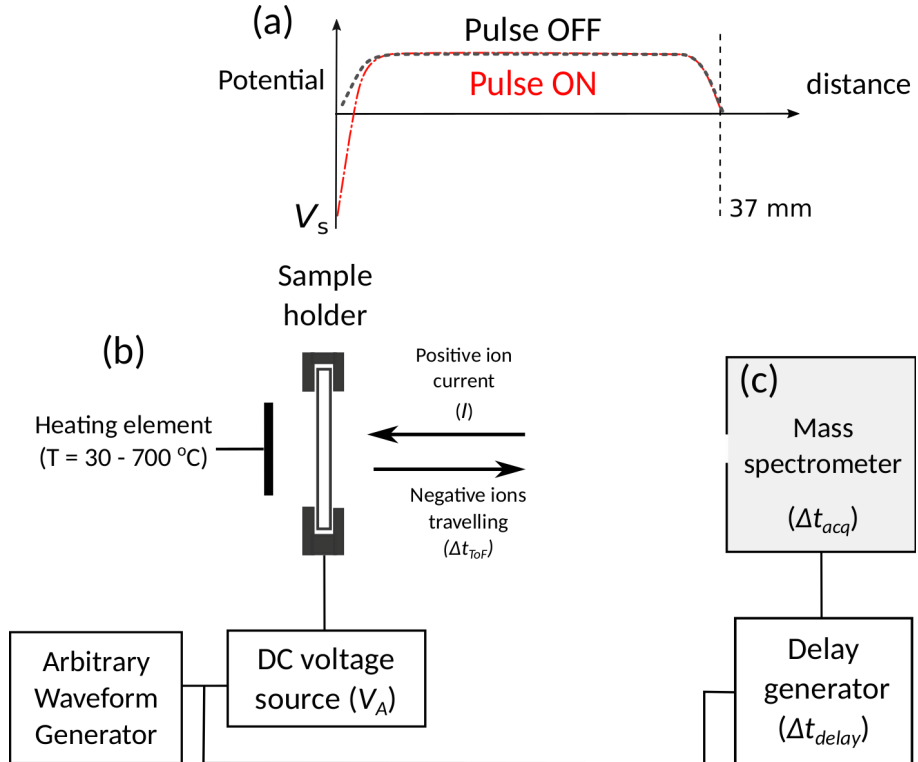}		\caption{Schematic of the sample holder and mass spectrometer timing setup. (a) Representative plasma potential profile between the sample holder and the mass spectrometer.
		(b) Sample holder with tungsten heating element and sample biasing electronics.
		(c) Mass spectrometer and delay generator for measurements of the negative ion energy distribution function (NIEDF) with respect to time within the sample bias pulse.}
		\label{fig:Experimental_diagram_split}
\end{figure}

A schematic of the sample holder and mass spectrometer, is shown in figure~\ref{fig:Experimental_diagram_split}.

Figure~\ref{fig:Experimental_diagram_split}~(a) shows a plasma potential profile that can be reasonably expected in the space between the sample surface and the mass spectrometer. When a bias (\textit{V\textsubscript{A}}) is applied to the sample using the DC voltage source as shown in figure~\ref{fig:Experimental_diagram_split}~(b), the voltage on the sample surface (\textit{V\textsubscript{S}}) decreases. This accelerates positive ions onto the sample surface. Negative ions that are produced due to the positive ion bombardment are accelerated by the negative bias such that they then cross the 37~mm gap separating the sample surface and the mass spectrometer. The mass spectrometer is then used to measure an NIEDF~\cite{Schiesko2008}.

The relatively low gas pressure of the plasma limits ion-neutral collisions within the diffusion chamber~\cite{AAhmad12013}. Any collisions that occur between the negative ions produced at the sample surface and the background gas are assumed to cause detachment, resulting in the destruction of the negative ion~\cite{Hemsworth2005, Schiesko2008}. The negative ions formed through volume production processes are prevented from entering the mass spectrometer due to the plasma potential in front of the mass spectrometer orifice. This means that in the absence of a negative bias, negative ions will not be not detected.

It is useful to also consider other potential sources of negative ions that may be detected by the mass spectrometer e.g. production in the sheath adjacent to the sample surface and production at the sample surface through the recombination of atomic deuterium to form an excited deuterium molecule. In the first case, the formation of negative ions in the sheath would have to be through dissociative attachment. This requires an electron to impact an excited molecule of deuterium. As the sheath is electron deficient, this is considered to be an unlikely source of negative ions. In the second case, the production of negative ions on the sample surface by neutral species would result in a peak in the NIEDFs at 0~eV as the negative ions would effectively be formed at rest. This is not observed in the experiments, which suggests that it has a negligible contribution to the NIEDF. Previous work has also investigated the production of negative ions from samples in a similar experimental setup using SRIM simulations\cite{Cartry2012, AAhmad12013}.  This demonstrated that the main contribution of negative ions from sample surfaces is from positive ion bombardment.

To investigate the influence of positive ion energy and negative ion yield, two positive ion bombardment energies are considered. A ``high-energy'' bombardment regime using an applied bias of -130~V, and a ``low-energy'' regime using a bias of -20~V. Bombarding positive ions are understood to dissociate into their constituent components during their interaction with the sample surface\cite{Cartry12017, Babkina2005a}. The plasma is primarily composed of D\textsubscript3\textsuperscript{+} ions, resulting in a positive ion energy bombardment of 44~eV/nucleon and 8~eV/nucleon for a negative bias of -130~V and -20~V respectively~\cite{AAhmad12013}. This means that the main contribution to negative ion formation is from D\textsubscript3\textsuperscript{+} ions and a smaller contribution is expected from D\textsubscript2\textsuperscript{+} ions and D\textsuperscript{+} ions. The Langmuir probe measurements in section~\ref{sec:langmuir_probe_motivation} suggest that the sheath width is smaller than ion-neutral mean free path. It is therefore reasonable to suggest that the positive ions undergo minimal collisions when passing through the sheath.

As described above, the positive ions bombard the sample surface to produce negative ions. These are then detected by the mass spectrometer, producing an NIEDF. Once the measurement has been completed, the NIEDF can be shifted to account for the energy the negative ions possessed when they were formed. This is possible because the total negative ion energy, \textit{E}, is conserved:~\cite{AAhmad12013}

\begin{equation}\label{eqn:energy_ion}
E = E_{\text{\textit{k}}} - eV_{\text{\textit{S}}}
\end{equation}

where \textit{E}\textsubscript{\textit{k}} is the kinetic energy of the negative ion when it was formed and \textit{V}\textsubscript{\textit{S}} is the voltage on the surface of the sample.The NIEDFs in this article are given in terms of~\textit{E}\textsubscript{\textit{k}}. 

In this work, both a pulsed bias and a continuous bias is utilised. When using a continuous bias in combination with a conductive sample, it is reasonable to expect that \textit{V\textsubscript{A} is equivalent to \textit{V\textsubscript{S}} and that the sheath in front of the sample is planar because the sample holder is much larger than the area of the sample from which negative ions are emitted compared to the sheath width. When a pulsed bias is applied to a non-conductive sample\cite{Achkasov2019}}, \textit{V\textsubscript{S}} becomes time dependent as a result of a build of charge on the sample surface\cite{Achkasov2019}. Therefore equation~\ref{eqn:energy_ion} is rewritten as:

\begin{equation}\label{eqn:energy_ion_pulse}
E = E_{\text{\textit{k}}} - eV_{\text{\textit{S}}}(t)
\end{equation}

\textit{V\textsubscript{\textit{S}(\textit{t})} on a non-conductive sample is calculated by considering the system as a capacitor:}

\begin{equation}\label{eqn:Vs}
V_{S}(t) = \frac{Q(t)}{C} + V_{A} + V_{f}
\end{equation}

where \textit{Q} is the charge build up on the surface of the sample due to positive ion bombardment, \textit{V\textsubscript{f}} is the floating potential on the sample before the application of the pulse (\textit{V\textsubscript{f}}~~=~0~V as measured with the Langmuir probe), and \textit{C} is the capacitance of the sample surface in contact with the plasma.


\subsection{\small{Negative ion yield using pulsed and continuous substrate biasing}}\label{sec:negative_ion_yield_calculation}

The negative ion yield is used as a means to compare the production of negative ions between samples, applied voltages and biasing techniques. This is defined as:

\begin{equation}\label{eqn:negative_ion_yield}
\alpha = \frac{1}{I} \int N_{D^{\text{-}}}(E)~\text{d}{E}
\end{equation}

where \textit{N\textsubscript{D\textsuperscript{-}}} is the number of negative ion counts detected by the mass spectrometer, which are integrated with respect to incident ion energy, \textit{E}, and \textit{I} is the positive ion current to the sample. As only a proportion of the negative ions that are produced by the sample surface are collected by the mass spectrometer, the negative ion yield is a relative measurement for comparing between samples. 

\subsubsection{\footnotesize{Electrical conditions for pulsed sample biasing}}\label{sec:pulsed_bias_timing}

\begin{figure}[H]
	\centering
		\includegraphics[]{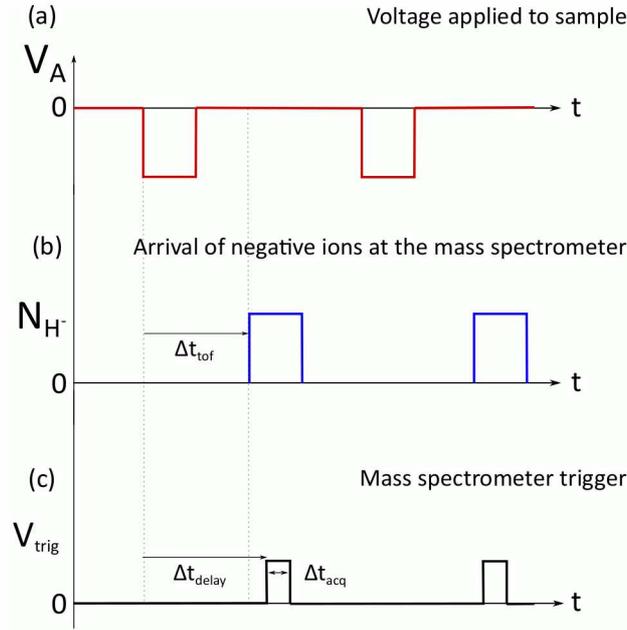}
		\caption{Timing diagram for an NIEDF measurement for pulsed bias operation. (a) Voltage applied to the sample (\textit{V\textsubscript{A}}) generates an electric field that causes positive ion bombardment and accelerates newly created negative ions towards the mass spectrometer (b) Negative ions travel from the sample to the mass spectrometer detector, arriving after a flight time, \textit{$\Delta$t}\textsubscript{tof}. (c) Measurement trigger voltage pulse (\textit{V}\textsubscript{trig}) sent from delay generator, shown in figure~\ref{fig:Experimental_diagram}, initiates a measurement of length \textit{$\Delta$t}\textsubscript{acq} by the mass spectrometer.}
    \label{fig:Change_in_surface_pot}
\end{figure}

As described in section~\ref{sec:mass_spec}, a surface bias is required to accelerate positive ions towards the sample surface. Pulsed biasing was undertaken to generate the necessary electric field at the surface of non-conductive samples, as shown in figure~\ref{fig:Change_in_surface_pot}.

In figure~\ref{fig:Change_in_surface_pot}~(a), a negative bias is applied to the sample (\textit{V\textsubscript{A}}) through the use of a waveform generator and DC voltage source, shown previously in figure~\ref{fig:Experimental_diagram_split}~(b). The negative voltage on the sample surface accelerates positive ions towards the sample and negative ions towards the mass spectrometer. The time of flight (\textit{$\Delta$t}\textsubscript{tof}) for these negative ions is shown in figure~\ref{fig:Change_in_surface_pot}~(b) and was measured to be approximately 14.5~\si{\micro\second} or 15.5~\si{\micro\second} when operating with ``high'' and ``low'' energy positive ion bombardment, respectively. The delay generator, shown in figure~\ref{fig:Experimental_diagram_split}~(c) sends a trigger to the mass spectrometer after time \textit{$\Delta$t}\textsubscript{delay}, for a duration corresponding to~\textit{$\Delta$t}\textsubscript{acq}. The time \textit{$\Delta$t}\textsubscript{delay} accounts for the time of flight (\textit{$\Delta$t}\textsubscript{tof}) of the negative ions from the samples to the mass spectrometer detector and is adjusted so that a NIEDF measurement, with a duration of \textit{$\Delta$t}\textsubscript{acq}, is acquired within the negative ion pulse arriving at the mass spectrometer as shown in figure~\ref{fig:Change_in_surface_pot}~(c).

The creation of negative ions on an insulating surface through positive ion bombardment causes the accumulation of positive charge on the sample surface. This implies that, at \textit{t}~=~0~\si{\micro\second}, \textit{Q}~=~0~C which means that \textit{V\textsubscript{S}} is equivalent to \textit{V\textsubscript{A}} as shown in equation~ref{eqn:Vs}. This corresponds to the case where a conductive sample is employed~\cite{Achkasov2019}. 

With reference to equation \ref{eqn:Vs}, the build up of positive charge changes the voltage on the surface, \textit{V\textsubscript{S}(t)}, over the duration of the pulse, which alters the energy that the negative ions possess when they are measured by the mass spectrometer ~\cite{Achkasov2019}. If the duty cycle is too high, the positive charge that accumulates during the ``on'' phase of the pulse will not dissipate before the next pulse~\cite{Achkasov2019}. Therefore a sufficiently small duty cycle is utilised to enable sufficient time for the surface potential of the sample to return to the floating potential via a recombination of positive charge on the sample surface with the incident electron flux.

\begin{figure}[H]
	\centering
		\includegraphics[]{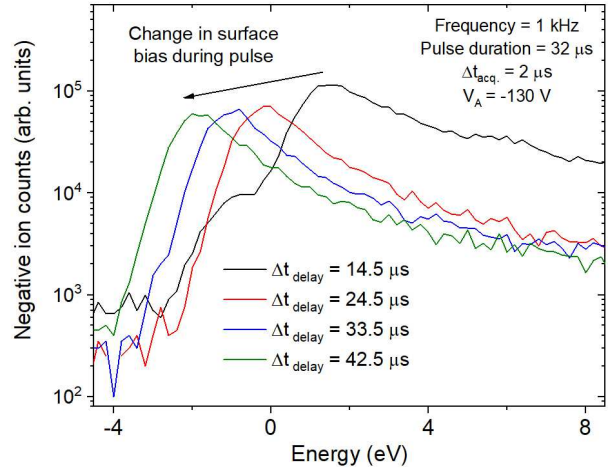}
		\caption{NIEDFs measured for pulsed-bias operation of an un-heated MCD sample with respect to the delay time between the application of the bias voltage and negative-ion measurement trigger, \textit{$\Delta$t}\textsubscript{delay}. Bias voltages are applied at 1~kHz with a duration of 32~\si{\micro\second} and applied voltage of -130~V. Low pressure deuterium plasma is operated at 2~Pa and 130~W.}
    \label{fig:voltage_shift}
\end{figure}

A series of NIEDFs were taken with increasing \textit{$\Delta$t}\textsubscript{delay} where the sample was biased using a square waveform pulse at a frequency of 1~kHz, an amplitude of negative -130~V and duration of 32~\si{\micro\second} (3.2\% duty cycle). The NIEDFs, shown in figure~\ref{fig:voltage_shift}, are generated using an insulating un-heated MCD sample with measurement duration ($\Delta$\textit{t}\textsubscript{acq}) of 2~\si{\micro\second}.

In figure~4, increasing $\Delta$t\textsubscript{delay} results in a shift in the peak of the NIEDF towards lower ion energies. This is because as $\Delta$t\textsubscript{delay} increases, the surface bias \textit{V\textsubscript{S}} starts to increase due to a build up of positive charge on the non-conductive surface. Within a single pulse, at \textit{t}~=~0~\si{\micro\second}, the charge built up on the sample surface, \textit{Q} from equation~\ref{eqn:Vs}, will be zero. This means that the applied voltage, \textit{V\textsubscript{A}} will be approximately equal to \textit{V\textsubscript{S}}.    
In the case where a sample is conductive, equation~\ref{eqn:energy_ion} can be used in the place of equation~\ref{eqn:energy_ion_pulse} and \textit{V\textsubscript{S}} can be substituted with \textit{V\textsubscript{A}} so that the NIEDFs of a conductive sample can be presented in terms of \textit{E\textsubscript{k}} by rearranging equation~\ref{eqn:energy_ion}\cite{Schiesko2008}. In the case where a sample is non conductive as shown in figure~\ref{fig:voltage_shift} the difference between \textit{V\textsubscript{A}} and \textit{V\textsubscript{S}} will increase over time. This results in the observed shift in figure~\ref{fig:voltage_shift} the negative ion energy peak as $\Delta$t\textsubscript{delay} is increased~\cite{Achkasov2019}.

Changes in \textit{V\textsubscript{S}} during $\Delta$t\textsubscript{acq} resulting in a ``smeared'' NIEDF are avoided by utilising a short $\Delta$t\textsubscript{acq}. The data in figure~\ref{fig:voltage_shift} demonstrates that a change in the surface voltage of 0.1~\si{\volt\micro\second}\textsuperscript{-1} can be expected for insulating samples when using a duty cycle of 3.2\%. $\Delta$t\textsubscript{acq} was chosen to be 2~\si{\micro\second} due to this being the longest time over which a change in the surface voltage on an insulating sample used in this study would be below the resolution of the mass spectrometer, thus maximising the negative ion signal, while minimising the shift in the surface voltage during the acquisition. 

The pulse condition that was chosen for this work was a square waveform with a 5~kHz frequency and 6~\si{\micro\second} duration (duty cycle 3\si{\percent}), which was selected to minimise the accumulation of positive charge on the surface of an insulating sample and therefore enable a reasonable comparison with samples that are conductive.

\subsection{\small{Measurement of the positive ion current}}\label{sec:positive_ion_current}

To measure the positive ion current, an electrically isolated copper electrode was inserted into the sample holder in the place of a sample. This electrode was isolated from the rest of the sample holder through the use of Kapton tape. The molybdenum bracket, used to affix samples to the sample holder, acted as a guard ring on the electrode to reduce edge effects interfering with the measurement of the positive ion current. The electrode and sample holder were electrically connected in parallel to reduce differences in sheath expansion from affecting the positive ion current measurement. 

\subsubsection{\footnotesize{Continuous bias operation}}\label{sec:continuous_bias_positive_ion_current}

The current drawn from the plasma was measured using an ammeter. This was done without any heating applied to the electrode to reduce the chance of overheating which could damage the Kapton tape. Variations in the current to the sample due to increases in its temperature are expected to be approximately 5~\%, as determined from separate measurements of the current to the sample holder during the experiments~\cite{Smith2020}. 

The measurement of the current to the sample at an applied voltage of -20~V was 14.5~\si{\micro\ampere}, whilst at -130~V the current was 18~\si{\micro\ampere}. The expected sheath width is approximately 1~mm or 2~mm, as determined using Langmuir probe measurements described in section~\ref{sec:langmuir_probe_motivation} for a -20~V and -130~V bias respectively. This is much smaller than the size of the sample holder, which is approximately a rectangle of side lengths of 3~cm. Therefore the sheath is expected to be approximately planar across the surface of the sample. 

\subsubsection{\footnotesize{Pulsed bias operation}}\label{sec:pulsed_bias_current}

\begin{figure}[H]
	\centering
		\includegraphics[]{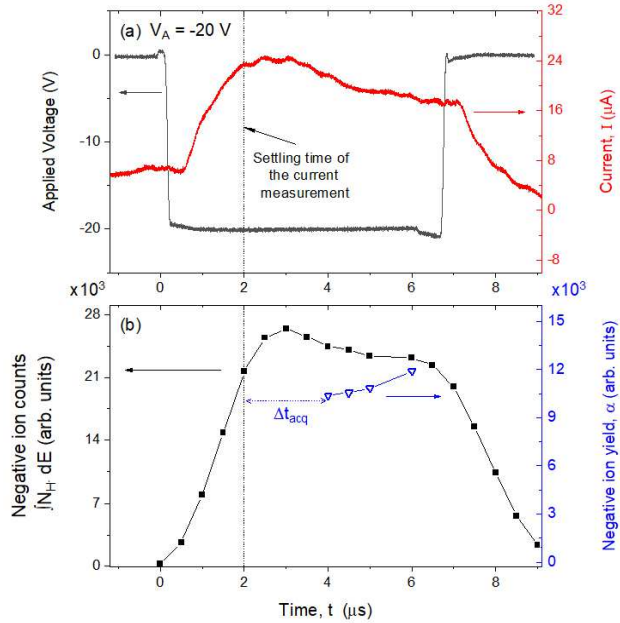}
		\caption{ (a) Voltage applied to sample holder and copper electrode with corresponding positive ion current. (b) Negative ion counts measured with incrementally increasing $\Delta$t\textsubscript{delay} for a $\Delta$t\textsubscript{acq} of 2~\si{\micro\second}. In (b), $\Delta$t\textsubscript{acq} is shown for the first data point of the negative ion yield measurement at t~=~4~\si{\micro\second}. Low pressure deuterium plasma is operated at 2 Pa and 130 W. Solid lines have been added to guide the eye.}
    \label{fig:Data_for_the_pulse}
\end{figure}

In pulsed bias operation, the instantaneous positive ion current was determined using the same setup as described in section \ref{sec:positive_ion_current} for the continuous bias, with a copper electrode inserted into the sample holder in place of a sample. A current-voltage converter, based upon a trans-impedance amplifier, was custom built to measure the sample current with an oscilloscope (Teledyne LeCroy Wavesurfer 4mXs-B). The current within a pulse was first measured with the plasma on, and then with the plasma off to account for leakage currents that are present due to the parasitic capacitance of the cables. The effective sample current was obtained by subtracting plasma off measurement from the plasma on measurement. Representative measurements of the applied voltage and positive ion current to the sample, and corresponding time-resolved negative ion yield, are shown in figure~\ref{fig:Data_for_the_pulse}.

Figure~\ref{fig:Data_for_the_pulse}~(a) shows the time-resolved applied voltage and positive ion current. The voltage that is applied to the sample is measured to be -20~V with little observed change during the application of the bias.


The settling time of the electrical measurement setup is determined to be $\sim$2~\si{\micro\second} by replacing the electrode with a resistor outside of the plasma chamber. This means that current measured between the application of the bias in figure~\ref{fig:Data_for_the_pulse} at 0~\si{\micro\second} and 2~\si{\micro\second} is considered to be unreliable due to the rapid change of the system at during this interval. Therefore, the period of time where the positive ion current can be reliably interpreted is between \textit{t}~=~2~\si{\micro\second} and the end of the pulse at \textit{t}~=~6.7~\si{\micro\second}, as shown in figure~\ref{fig:Data_for_the_pulse}~(a).  During this interval, the current is observed to first peak at 25~\si{\micro\ampere} and then decrease steadily to 17~\si{\micro\ampere}. It is observed that the current is higher during the application of the pulsed bias compared to the use of the continuous bias. This is consistent with observations that the current decreases over a time to a similar value observed when using a continuous sample bias. It is therefore reasonable to suggest that the relatively high current observed at the beginning of the pulse, i.e. at 2~\si{\micro\second} in figure~\ref{fig:Data_for_the_pulse}~(a), is consistent with rapid perturbation and stronger disturbance of the plasma close to the sample holder when the applied voltage switches.




Figure~\ref{fig:Data_for_the_pulse}~(b) shows the time-resolved negative ion counts detected during the bias pulse and the corresponding negative-ion yield, as determined with equation~\ref{eqn:negative_ion_yield}, for MCD at 550$^{\circ}$C. This temperature is chosen because in previous work a temperature of 550$^{\circ}$C produced the largest quantity of negative ions, whilst also ensuring that the sample is fully conductive~\cite{Smith2020}. 
From 0~\si{\micro\second}, the negative ion counts increases from a negligible amount of negative ions to a peak at 1~\si{\micro\second}. This roughly aligns with the measured peak in the positive ion current, shown in figure~\ref{fig:Data_for_the_pulse}~(a). After this peak the counts decrease by \num{3e3} over the duration of the pulse, before decaying to zero rapidly at the end of the pulse. Negative ions are measured during an acquisition window of $\Delta$t\textsubscript{acq}~ which in this study is set to 2~\si{\micro\second}. It is observed in figure~\ref{fig:Data_for_the_pulse}~(a) that the current changes by approximately 5~\si{\micro\ampere} over the course of $\Delta$\textit{t}\textsubscript{acq}. 



To compare the relative negative ion yield between samples, a consistent temporal position in the negative ion pulse at \textit{t}~=~4~\si{\micro\second} encompassing a $\Delta$t\textsubscript{acq} window between 2~\si{\micro\second} and 4~\si{\micro\second} is chosen, as shown in figure~\ref{fig:Data_for_the_pulse}~(b). This period was chosen as it is the earliest point in time that the current can be determined, thus minimising the build up of positive ions on the surface of the samples.

\section{\large{Negative ion yield: comparison of MCD and MCNDD for pulsed and continuous biases }}\label{sec:pulsedbiascomparison}

\begin{figure*}
	\centering
		\includegraphics[]{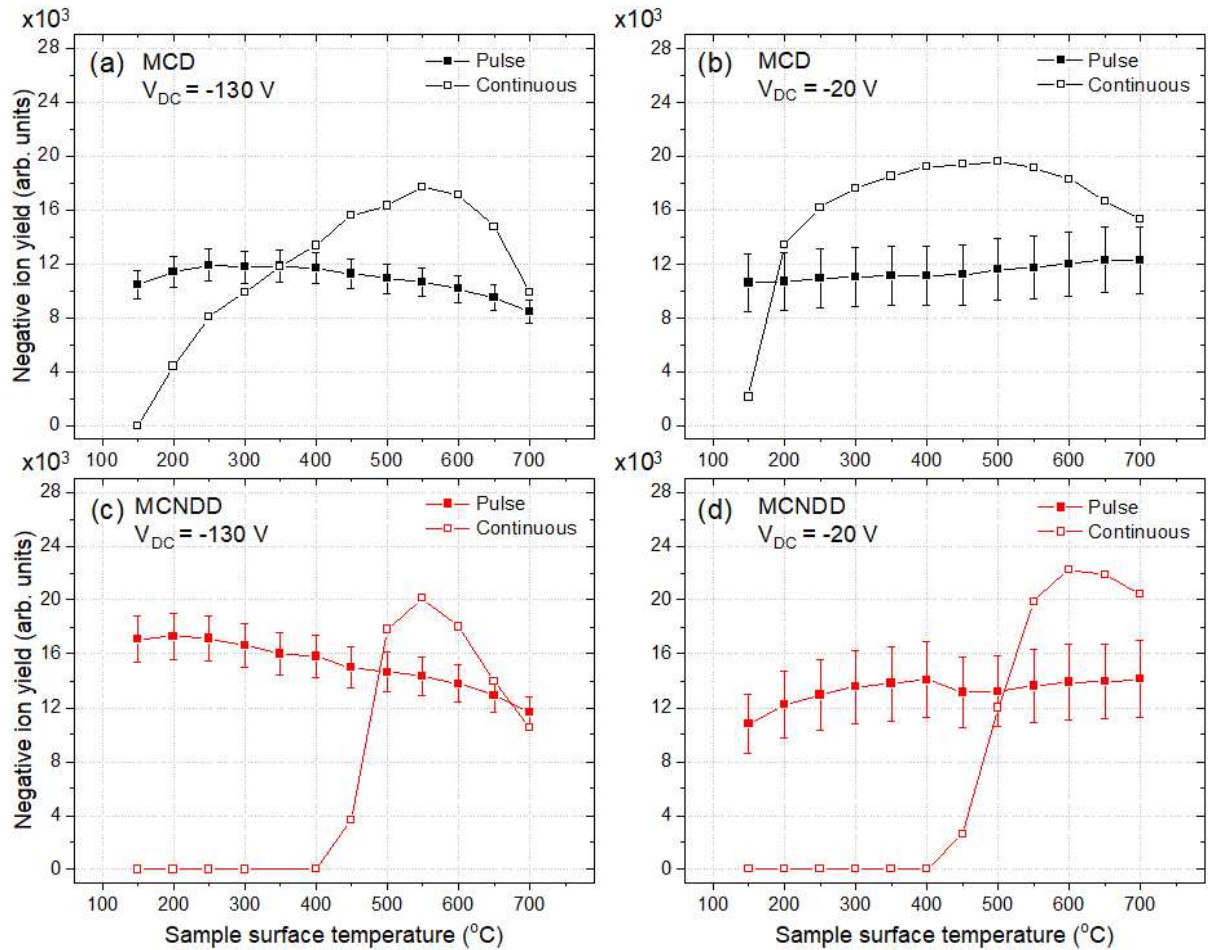}
		\caption{Negative ion yield with respect to temperature of the sample surface for continuous and pulsed biased operation. (a) MCD with a bias of $-$130~V  (b) MCD with a bias of $-$20~V (c) MCNDD with a bias of $-$130~V (d) MCNDD with a bias of $-$20~V. Uncertainty bars of 10\% and 20\% have been included for the pulsed bias yield measurements at -130~V and -20~V respectively to account for the uncertainty in the current measurements in each respective case. Low pressure deuterium plasma is operated at 2~Pa and 130~W. Solid lines are included as visual guide.}
    \label{fig:Pulsed_comparison_yield_20-130V_Diamond_4panelfigure}
\end{figure*}

Figure~\ref{fig:Pulsed_comparison_yield_20-130V_Diamond_4panelfigure} shows the negative ion yield measured using MCD and MCNDD for ``high-energy'' and ``low-energy'' positive ion bombardment conditions in continuous and pulsed sample bias operation.

Uncertainty bars have been included in the pulsed bias case to account for the change in the current during the 2~\si{\micro\second} acquisition interval, $\Delta$t\textsubscript{acq}.

\subsection{\small{Negative ion yield for continuous bias operation}}\label{sec:continuous_bias}

In figure~\ref{fig:Pulsed_comparison_yield_20-130V_Diamond_4panelfigure}~(a) the negative ion yield for MCD is observed to increase up to \num{17.7e3} as the sample temperature is increased from 150$^{\circ}$C to 550$^{\circ}$C when using a -130~V continuous sample bias. This is then followed by a decrease in the yield from \num{17.7e3} to \num{9.8e3} as the temperature is increased from 550$^{\circ}$C to 700$^{\circ}$C. 

In figure~\ref{fig:Pulsed_comparison_yield_20-130V_Diamond_4panelfigure}~(b) when using a -20~V bias, the trend of increasing yield with increasing sample temperature is similar to that observed in figure~\ref{fig:Pulsed_comparison_yield_20-130V_Diamond_4panelfigure}~(a), it increases from \num{2e3} to \num{20e3} as the temperature increases from 150$^{\circ}$C to 500$^{\circ}$C, and then decreases to a yield of \num{15e3} from a sample temperature of 500$^{\circ}$C to 700$^{\circ}$C. 


The negative ion yield from MCNDD samples is shown in figures~\ref{fig:Pulsed_comparison_yield_20-130V_Diamond_4panelfigure}~(c) and (d). When using a continuous sample bias, at sample temperatures below 400$^{\circ}$C the yield from MCNDD cannot be measured due to the poor conductivity of the sample. Once the MCNDD sample temperature reaches 400$^{\circ}$C, negative ions are measured and a peak in the negative ion yield of \num{20e3} at 550$^{\circ}$C is observed in figure~\ref{fig:Pulsed_comparison_yield_20-130V_Diamond_4panelfigure}~(c) when using a sample bias of -130~V. Similarly, a negative ion yield of \num{22e3} at 600$^{\circ}$C is observed in figure~\ref{fig:Pulsed_comparison_yield_20-130V_Diamond_4panelfigure}~(d) when using a sample bias of -20~V. The changes in the negative ion yield when using a continuous sample bias for MCNDD and MCD at temperatures above 550$^{\circ}$C are similar to results of previous work\cite{Smith2020}. The peak negative ion yield from MCNDD is about 10\% higher than MCD for a -20V bias, which is also consistent with previous work\cite{Smith2020}.

Differences between the samples and biasing methods that can be observed in figure~\ref{fig:Pulsed_comparison_yield_20-130V_Diamond_4panelfigure} can be understood by analysing the corresponding NIEDFs, which are used to determine the negative ion counts coming from the samples. The distribution of negative ions in the NIEDFs have previously been shown to be useful for determining the negative ion production mechanisms involved when using carbon samples\cite{AAhmad12013}. There are two production processes that are considered to be responsible for negative ion formation from carbon: backscattering and sputtering. Backscattering produces negative ions when an incoming positive ion is reflected off the deuterated carbon lattice of the samples and during the collision the positive ion captures two electrons\cite{Schiesko2010c}. As distinct from this, sputtering relies on the ejection of adsorbed deuterium from the carbon lattice as a negative ion\cite{Schiesko2010c}. 

\begin{figure*}
	\centering
		\includegraphics[]{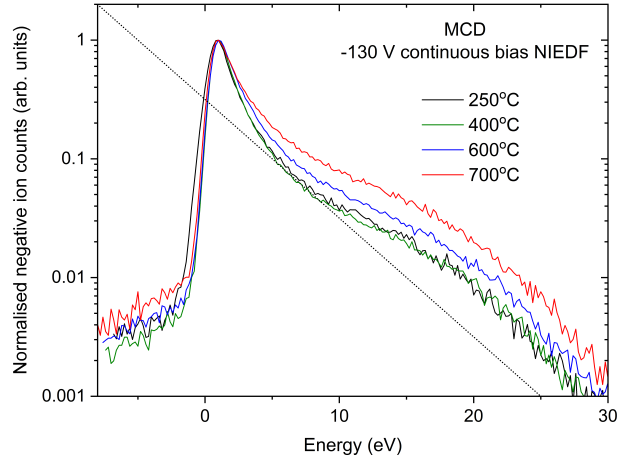}
		\caption{NIEDF measurements for MCD at sample temperatures of 250$^{\circ}$C, 400$^{\circ}$C, 600$^{\circ}$C and 700$^{\circ}$C using a bias of -130~V, applied continuously. Dotted reference line added to guide the eye. Low pressure deuterium plasma is operated at 2 Pa and 130 W.}
    \label{fig:Comparison_0ppm_temp_NIEDF}
\end{figure*}

Of the two production processes, sputtering is temperature dependent because it relies on the presence of adsorbed deuterium on and within the sub-surface lattice of the diamond samples. Increasing the sample temperature reduces the adsorption of deuterium and encourages out-gassing into the diamond lattice, thus reducing the amount of sputtering that can occur\cite{Schiesko2010c, Ahmad2014, AAhmad12013}. 
Changes in the proportion of negative ions produced by the two production processes can be observed by considering the NIEDFs produced by samples at different temperatures. Figure~\ref{fig:Comparison_0ppm_temp_NIEDF} shows the NIEDFs for MCD at temperatures of 250$^{\circ}$C, 400$^{\circ}$C, 600$^{\circ}$C and 700$^{\circ}$C. 

In figure~\ref{fig:Comparison_0ppm_temp_NIEDF}, the NIEDFs shown have been normalised to the peak negative ion counts. In previous work, it was observed that the proportion of negative ions produced through sputtering processes was higher at lower negative ion energies than at higher energies\cite{Cartry2012}. This means that, due to the normalisation of the NIEDFs, a change in the tail height can be used to infer the relative proportion of backscattering compared to sputtering, which in turn can be used to infer the amount of sub-surface deuterium is in the samples\cite{Cartry2012, Ahmad2014, Dubois2016}. It is observed that the tail of the NIEDFs in figure~\ref{fig:Comparison_0ppm_temp_NIEDF} rises as the sample temperature is increased. When combined with observations in figure~\ref{fig:Pulsed_comparison_yield_20-130V_Diamond_4panelfigure}~(a)~and~(c) that show a decrease in the negative ion yield at temperatures above 550$^{\circ}$C for both MCD and MCNDD, it demonstrates that a reduction in the adsorbed deuterium within the sample decreases the negative ion yield from the samples. This observation is consistent with previous work\cite{Smith2020, Ahmad2014, Cartry2012}. A similar comparison cannot be carried out using the NIEDFs produced using a -20~V bias as the NIEDFs in this case do not have a high energy tail. However, it is reasonable to suggest that a similar process occurs when using a -20~V bias as the negative ion yield decreases in a similar manner when using a continuous sample bias past a sample temperature of 500$^{\circ}$C and 600$^{\circ}$C for MCD and MCNDD respectively.


As shown in figure~\ref{fig:Pulsed_comparison_yield_20-130V_Diamond_4panelfigure}~(a)~and~(b), when applying a continuous bias to MCD, the negative ion yield increases as the temperature of the sample is increased from 150$^{\circ}$C to 550$^{\circ}$C. This is consistent with previous work and has been attributed to a coupled process of defect formation, where the diamond sp3 bonds are turned to sp2 bonds by the bombarding positive ions from the plasma, and then the preferential etching away of these newly created sp2 bonds\cite{Ahmad2014, Kogut2019, Roth1996}. Surfaces that are composed mostly of sp2 bonds have previously been associated with a reduction in the negative ion yield. Therefore the creation of these bonds has previously been expected to decrease negative ion yield if too many exist on a diamond surface\cite{Ahmad2014}. By increasing the temperature of the samples, the rate at which the sp2 bonds are etched away will be increased resulting in a reduction in the number of defects on the surface at higher sample temperatures, and thus a higher negative ion yield from samples that are heated\cite{Ahmad2014, Kogut2019, Roth1996}. This process is observed to cause a peak in the negative ion yield at a temperature of approximately 550$^{\circ}$C for MCD, before the higher temperature of the samples causes a reduction in the sputtering contribution.

Previous work has demonstrated using ex-situ Raman spectroscopy that exposure of samples of diamond to positive ion bombardment is associated with an increase in the amount of defects on the sample surface\cite{Kogut2019, Ahmad2014}. A similar process of defect formation can reasonably be expected in this work as the trends observed when heating the samples under a continuous positive ion bombardment are the same as that observed in previous work\cite{Smith2020, Ahmad2014, Kogut2019}. These trends are observed in figure~\ref{fig:Pulsed_comparison_yield_20-130V_Diamond_4panelfigure}~(a)~and~(b) when comparing the negative ion yield between a sample bias of -130~V and -20~V, respectively. A ``high-energy'' positive ion bombardment will produce more defects. At lower sample temperatures, when using a``high-energy'' positive ion bombardment, the defects are not etched away and so a lower negative ion yield is observed due to a larger proportion of sp2 bonds. Increasing the temperature of the sample will increase the negative ion yield as the defect bonds are etched away more rapidly. This results in a lower negative ion yield from MCD at temperatures between 150$^{\circ}$C and 550$^{\circ}$C when using a ``high-energy''~(\textit{V\textsubscript{A}}~=~-130~V) positive ion bombardment compared to a  ``low-energy''~(\textit{V\textsubscript{A}}~=~-20~V) positive ion bombardment across the same temperature range. This also means that a higher peak negative ion yield for a low energy positive ion bombardment can be expected, as the number of defects is reduced when using a ``low-energy'' positive ion bombardment. This is observed in figure~\ref{fig:Pulsed_comparison_yield_20-130V_Diamond_4panelfigure}~(b) at 500$^{\circ}$C which is higher that the peak in the negative ion yield at 550$^{\circ}$C in figure~\ref{fig:Pulsed_comparison_yield_20-130V_Diamond_4panelfigure}~(a). 

In figure~\ref{fig:Pulsed_comparison_yield_20-130V_Diamond_4panelfigure}~(c)~and~(d), the negative ion yield from MCNDD is observed to decrease with respect to sample temperature at a similar rate to MCD at applied biases of -130~V and -20~V. A similar comparison to MCD cannot be made at temperatures below 550$^{\circ}$C as addition of nitrogen doping lowers the conductivity of the diamond such that the negative ion yield cannot be measured at sample temperatures below 550$^{\circ}$C. 



\subsection{\small{Negative ion yield for pulsed bias operation}}

In figure~\ref{fig:Pulsed_comparison_yield_20-130V_Diamond_4panelfigure}~(a), the negative ion yield from MCD when using a pulsed sample bias of -130~V increases as the sample temperature is increased from 150$^{\circ}$C to 250$^{\circ}$C by \num{1.4e3} and then decreases as the sample temperature increases from 250$^{\circ}$C to 700$^{\circ}$C by \num{3.4e3}. This is unlike the trend observed in figure~\ref{fig:Pulsed_comparison_yield_20-130V_Diamond_4panelfigure}~(b), where the negative ion yield, when using a pulsed bias of -20~V, increases as sample temperature is increased from 150$^{\circ}$C to 700$^{\circ}$C by \num{1.7e3}.

The negative ion yield from MCNDD using a pulsed sample bias is shown in figures~\ref{fig:Pulsed_comparison_yield_20-130V_Diamond_4panelfigure}~(c)~and~(d) for -130~V and -20~V sample bias voltages, respectively. Using a pulsed bias, negative ions are produced at temperatures lower than 450$^{\circ}$C, compared to when operating with a continuous sample bias. In figure~\ref{fig:Pulsed_comparison_yield_20-130V_Diamond_4panelfigure}~(c), the negative ion yield is shown to decrease by \num{5.4e3} as the sample temperature is increased from 150$^{\circ}$C to 700$^{\circ}$C. In contrast to this, in figure~\ref{fig:Pulsed_comparison_yield_20-130V_Diamond_4panelfigure}~(d) it is observed that the negative ion yield increases by \num{3.4e3} as sample temperatures are increased from 150$^{\circ}$C to 400$^{\circ}$C when using a pulsed sample bias of -20~V. 

In figure~\ref{fig:Pulsed_comparison_yield_20-130V_Diamond_4panelfigure}~(a)~and~(b) when using a pulsed sample bias, it is observed that there is a comparatively small change in the negative ion yield as sample temperature is increased, which is in contrast to the large change in negative ion yield observed when using a continuous sample bias between sample temperatures of 150$^{\circ}$C to 500$^{\circ}$C. As described in section~\ref{sec:continuous_bias}, the negative ion yield from diamond is influenced by the number of sp2 defects formed by bombarding positive ions. As the sample temperature is increased,  sp2 defects are etched at an increased rate. This means that the negative ion yield from a sample is expected to increase as the sample temperature is increased, as is observed when using a continuous bias. When using a pulsed -130~V sample bias, as shown in figure~\ref{fig:Pulsed_comparison_yield_20-130V_Diamond_4panelfigure}~(a), the negative ion yield peaks at a lower sample temperature compared to when using a continuous sample bias, at 250$^{\circ}$C, and then decreases as the sample temperature is increased. When using a pulsed sample bias case, if little to no defects are formed on the sample surface, the influence of temperature on the negative ion yield due to preferential etching will be reduced and only a decrease in the sputtering contribution will be observed. As there is a comparatively small change in the negative ion yield observed when using a pulsed sample bias compared to a continuous sample bias at -130~V, it can be confirmed that the sample surface is being preserved when using this technique, and that the pulsed sample bias measurements are representative of an almost ``pristine'' sample surface\cite{Achkasov2019}. 

By using a pulsed sample bias, it is possible to compare the negative ion yield from the pristine surface states of MCNDD and MCD at temperatures between 150$^{\circ}$C and 700$^{\circ}$C, as well as generate and measure negative ions in spite of the low conductivity of MCNDD at temperatures below 400$^{\circ}$C. The negative ion yield from MCNDD is observed to be, on average, higher than MCD by 28\% at -130V and 14\% at -20V across the compared temperatures. This supports previous work that observed that negative ion yield from diamond is enhanced when nitrogen doping is added to diamond samples and suggests that the surface state of MCNDD is conducive to enhanced negative ion yield. 



\begin{figure*}
	\centering
		\includegraphics[]{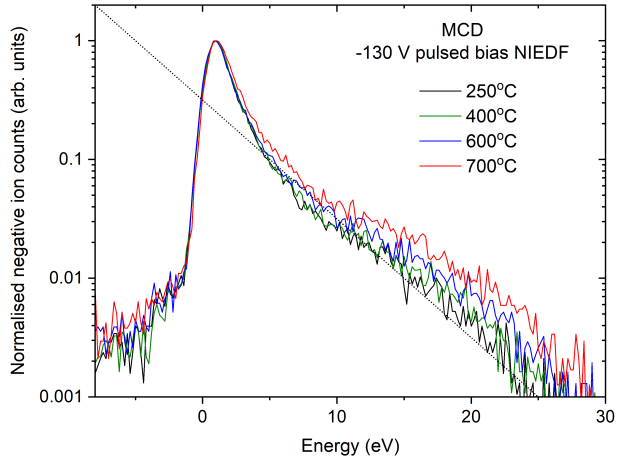}
		\caption{NIEDF measurements for MCD at sample temperatures of 250$^{\circ}$C, 400$^{\circ}$C, 600$^{\circ}$C and 700$^{\circ}$C using a pulsed bias of -130~V. Dotted reference line added to guide the eye. Low pressure deuterium plasma is operated at 2 Pa and 130 W.}
    \label{fig:Comparison_0ppm_temp_NIEDF_pulsed}
\end{figure*}

A series of NIEDFs when using a pulsed sample bias of -130~V and a MCD sample are shown in figure~\ref{fig:Comparison_0ppm_temp_NIEDF_pulsed} for sample temperatures of 250$^{\circ}$C, 400$^{\circ}$C, 600$^{\circ}$C and 700$^{\circ}$C. It can be observed in figure~\ref{fig:Comparison_0ppm_temp_NIEDF_pulsed} that there is a smaller change in the tail height when compared to figure~\ref{fig:Comparison_0ppm_temp_NIEDF}, suggesting a smaller change in the sputtering contribution to the negative ion yield as the sample temperature is increased over the same range of temperatures. 

The negative ion yield in figure~\ref{fig:Pulsed_comparison_yield_20-130V_Diamond_4panelfigure}~(a) when applying a pulsed bias is observed to decrease as the temperature of the MCD sample is increased from 200$^{\circ}$C to 700$^{\circ}$C. This observation is consistent with the observed changes in the tail height in figure~\ref{fig:Comparison_0ppm_temp_NIEDF_pulsed}. There is an increase in the tail height as the sample temperature is increased suggesting a similar decrease in the sputtering contribution that reduces the negative ion yield from the sample as sample temperature is increased.

Using the dotted reference lines in figures~\ref{fig:Comparison_0ppm_temp_NIEDF} and figure~\ref{fig:Comparison_0ppm_temp_NIEDF_pulsed} as a guide, we qualitatively compare the tail heights for pulsed and continuous sample biasing. By comparing the tail heights in these figures, it is observed that the amount of sputtering from a pulsed sample bias appears to be larger than that observed when using a continuous sample bias. This is consistent with previous work which suggested that a non-continuous flux of positive ions allows for the re-adsorbtion of deuterium lost through the sputtering process when the sample bias is being applied that would otherwise be depleted when utilising a continuous sample bias\cite{Achkasov2019}. It can also be observed that the increase in the tail height between a sample temperature of 600$^{\circ}$C and 700$^{\circ}$C, is much larger in figure~\ref{fig:Comparison_0ppm_temp_NIEDF} than in figure~\ref{fig:Comparison_0ppm_temp_NIEDF_pulsed}. This agrees with the larger decrease in the negative ion yield observed in figure~\ref{fig:Pulsed_comparison_yield_20-130V_Diamond_4panelfigure}~(a) between 600$^{\circ}$C and 700$^{\circ}$C when using a continuous sample bias compared to when using a pulsed sample bias.


For the same reason as the continuous sample biasing case, it is not possible to compare the NIEDFs for -20~V pulsed sample biasing to determine the sputtering contribution due to the lack of a high energy tail in the NIEDF. This means that the sputtering contribution from these samples is inferred by considering the NIEDFs using a -130~V bias and the trends observed in figure~\ref{fig:Pulsed_comparison_yield_20-130V_Diamond_4panelfigure}. As sample temperatures are increased, a decrease in the sputtering contribution can be expected as the adsorbed deuterium is out-gassed from the samples\cite{Ahmad2014}. This is observed to be the case for MCD and MCNDD in figure~\ref{fig:Pulsed_comparison_yield_20-130V_Diamond_4panelfigure}~(a)~and~(c) when using a -130~V pulsed sample bias. It is observed that the negative ion yield from MCD decreases as the sample temperature is increased from 150$^{\circ}$C to 700$^{\circ}$C by 28\% and by 41\% for MCNDD over a similar temperature scale. This is in contrast to the observations in  figure~\ref{fig:Pulsed_comparison_yield_20-130V_Diamond_4panelfigure}~(b)~and~(d) for MCD and MCNDD when using a -20~V pulsed sample bias. It is observed that the negative ion yield increases by 13\% over a similar temperature range for MCD and by 24\% for MCNDD. Previous work has suggested that the deuterium content of the samples is similar at similar sample temperatures\cite{Smith2020, Ahmad2014}. Therefore the increase in the negative ion yield as the sample temperature is increased suggests that when utilising both a ``low'' positive ion energy and a pulsed sample bias, the contribution of sputtering to the overall negative ion yield is reduced or even absent.

To explain the mechanism for a reduced sputtering contribution with a ``low-energy'' pulsed positive ion bombardment, it is worth considering previous work and differences in the negative ion yield at sample temperatures above 500$^{\circ}$C when using a continuous sample bias. Previous studies have determined that the threshold energy for the sputtering of hydrogen from carbon occurs at approximately 15~eV~\cite{Kogut2019}. As previously described in section~\ref{sec:langmuir_probe_motivation}, the sheath of the plasma in front of the samples can reasonably be expected to be collision-less. Therefore, the dominant positive ion, D\textsubscript3\textsuperscript{+}, upon impact with the sample surface will have an energy of 8~eV when using a -20~V bias\cite{Kogut2019}. This is below the calculated threshold for sputtering, suggesting that only a small amount of sputtering can occur when using a -20~V bias, where the lighter positive ions (D\textsubscript2\textsuperscript{+} and D\textsuperscript{+}) are able to exceed the threshold energy. An issue with this interpretation is that in figure~\ref{fig:Pulsed_comparison_yield_20-130V_Diamond_4panelfigure}~(b), when utilising a continuous sample bias of -20~V, the negative ion yield is observed to decrease as the sample temperature is increased from 500$^{\circ}$C to 700$^{\circ}$C which is consistent with a reduction in the sputtering contribution from the samples, despite the positive ion energy being below the threshold for sputtering to occur. The difference between these two cases is the type of sample biasing being used. Therefore, this observation suggests that the use of a -20~V pulsed sample bias reduces the contribution of sputtering to the negative ion yield and is possibly linked to defect formation caused by a continuous positive ion bombardment.

It is interesting to note that the negative ion yield observed when using a -20~V pulsed sample bias for MCD and MCNDD (figure~\ref{fig:Pulsed_comparison_yield_20-130V_Diamond_4panelfigure}~(b) and figure~\ref{fig:Pulsed_comparison_yield_20-130V_Diamond_4panelfigure}~(d), respectively) is similar at similar sample temperatures, compared to MCD and MCNDD when using a -130~V pulsed sample bias (figure~\ref{fig:Pulsed_comparison_yield_20-130V_Diamond_4panelfigure}~(a) and figure~\ref{fig:Pulsed_comparison_yield_20-130V_Diamond_4panelfigure}~(c), respectively). This suggests that the when using a pulsed sample bias, the surface state of the samples is similar despite the differences in the applied voltages. This further supports the argument that the pulsed bias preserves the surface state of the samples.


\subsection{\small{Comparing the negative ion yield between pulsed and continuous bias operation}}\label{sec:compared_bias}

In figure~\ref{fig:Pulsed_comparison_yield_20-130V_Diamond_4panelfigure}~(a)~and~(b), when using a -130~V and -20~V continuous sample bias respectively with MCD, the negative ion yield is observed to be almost 1.6 times higher than that observed from a similar sample when applying a -130~V pulsed sample bias at a sample temperature of 550$^{\circ}$C. Similarly for MCNDD, in figure~\ref{fig:Pulsed_comparison_yield_20-130V_Diamond_4panelfigure}~(c)~and~(d), the negative ion yield is observed to be 1.4 times higher when using a continuous bias compared to when applying a pulsed sample bias at a sample temperature of 550$^{\circ}$C. In addition to this, the differences between the pulsed and continuous sample biases appears to be consistent between all of the samples. This suggests that a similar mechanism for negative ion formation exists between each of the samples. 

Previous work observed that a pulsed bias produced a higher number of negative ion counts (\textit{N\textsubscript{D\textsuperscript{-}}}) from diamond samples and attributed this to a preserved surface state\cite{Achkasov2019}. This was based on the assumption that a pristine surface state composed primarily of sp3 is an ideal surface state for negative ion production from diamond, as discussed in section~\ref{sec:continuous_bias}. However, in this previous work a measurement of the positive ion current within the pulse was not available in order to calculate the negative ion yield and confirm that it is higher with a pristine surface\cite{Achkasov2019}. 

The observation of a lower negative ion yield when using a pulsed sample bias compared to a continuous one means it is reasonable to suggest that there is a change in the surface state of the samples that occurs as a result of the positive ion bombardment. This is because, as the pulse bias duty cycle is increased, the time the bias is applied to the sample will increase and will eventually be equivalent to a continuous sample bias, meaning the yield must increase as the pulse duration is increased. There are two proposed mechanisms that could change the surface state of the sample and therefore increase the negative ion yield. One is a change in the adsorbed deuterium which is responsible for sputtering from the samples, and the other is an increase in the number of sp2 bonds created on the sample surface that has been reduced by using a pulsed sample bias. 

\begin{figure}[H]
    \centering
    \includegraphics[]{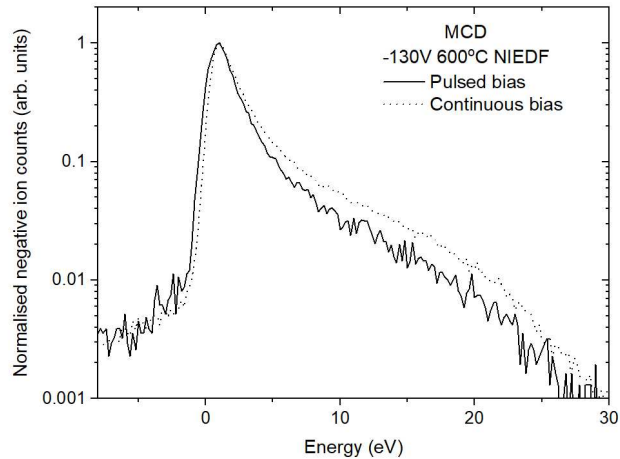}
    \caption{NIEDF measurements for MCD using pulsed or continuous bias of -130~V at 600$^{\circ}$C. Low pressure deuterium plasma is operated at 2~Pa and 130~W.}
    \label{fig:600C_NIEDF_comparison}
\end{figure}

As described previously, the deuterium content of the samples is observed to be higher when using a pulsed sample bias. This can be observed in figure~\ref{fig:600C_NIEDF_comparison}, which shows the NIEDFs for a continuous bias and a pulsed bias at 600$^{\circ}$C.

The tail height of the pulsed bias in figure~\ref{fig:600C_NIEDF_comparison} is lower than the continuous bias, which means that the negative ion yield contribution via the sputtering process is higher when using a pulsed bias. If the deuterium content of the sample is reduced when using a continuous bias, the negative ion yield can reasonably be expected to be reduced\cite{Ahmad2014}. However, if we consider the results in figure~\ref{fig:Pulsed_comparison_yield_20-130V_Diamond_4panelfigure}, we instead observe a increase in the negative ion yield when using a continuous sample bias compared to a pulsed sample bias. This therefore suggests that the sputtering contribution changing between a pulsed and continuous sample biasing is unlikely to be responsible for higher negative ion yield observed when using a continuous bias. 

The other suggested mechanism is that the negative ion yield is higher when using a continuous sample bias due to the bombardment of the sample surfaces creating sp2 defect bonds. This appears to be contrary to the conclusions of previous work. In previous work it has been shown that a surface composed entirely of sp2 bonds is detrimental to negative ion yield\cite{Ahmad2014}, meaning techniques to preserve the surface of diamond and minimise the number of sp2 bonds has been of interest\cite{Achkasov2019}. However, in this study the process of defect formation appears to be the only difference between the pulsed and continuous biasing technique that could be responsible for changes in the negative ion yield. This suggests that some degree of defect formation is important for increasing the negative ion yield from diamond. The authors believe that these observations suggest that an optimum surface state for negative ion yield possibly exists on diamond that is dependent on the pulsed bias frequency, doping, positive ion energy and temperature of the samples. This surface state will have an optimum ratio of sp2 and sp3 defects as a result of the interplay of the positive ion energy and flux to the samples. Further work to determine the optimum conditions is on going and in-situ time-resolved measurements of the sample surface state transitioning away from a ``pristine'' surface during plasma exposure remains the subject of future work.

\section{Conclusion}

In this study the surface production of negative ions from pulse-biased non-doped diamond (MCD) and nitrogen doped diamond (MCNDD) within a low-pressure deuterium plasma is investigated. The pulsed negative bias is applied in a square waveform pulse at 5~kHz, with a duty cycle of 3\% to preserve the surface of the samples and allow for measurement of negative ion yield in the absence of sample conductivity. The negative ion yield from MCNDD films when using a pulsed biased, determined via mass spectrometry and measurements of the positive-ion current to the sample, is observed to be higher than non-doped films at temperatures between 150$^{\circ}$C and 700$^{\circ}$C, confirming that nitrogen doping of diamond can be used to enhance negative ion yield when the surface state of the diamond is preserved. The pulsed bias has also been shown to have a lower peak negative ion yield compared to a continuous bias, which suggests that there exists an optimum sp\textsuperscript{2} to sp\textsuperscript{3} ratio for diamond sample surfaces to maximise negative ion yield.

\section{\large{Acknowledgements}}

The authors would like to acknowledge the experimental support of Jean Bernard Faure and the rest of the PIIM group. This work has been carried out within the framework of the French Federation for Magnetic Fusion Studies (FR-FCM) and of the EUROfusion consortium, and has received funding from the Euratom research and training programme 2014-2018 and 2019-2020 under grant agreement No.~633053. The views and opinions expressed herein do not necessarily reflect those of the European Commission. Financial support was received from the French Research Agency (ANR) under grant 13-BS09-0017 H INDEX TRIPLED. The financial support of the EPSRC Centre for Doctoral Training in fusion energy is gratefully acknowledged under financial code EP/L01663X/1. CGI (Commissariat \'a l'Investissement d'Avenir) is gratefully acknowledged for its financial support through Labex SEAM (Science and Engineering for Advanced Materials and devices) (No. ANR 11 LABX 086, IDEX 05 02).

\section{\Large{References}}   
    
\renewcommand\refname{}{\vskip -1cm}

\begin{footnotesize}
\bibliography{Bibliography_paper2.bib}
\bibliographystyle{unsrtnat}
\end{footnotesize}
\end{document}